\documentclass[aps, prb, reprint, lengthcheck, footinbib, showpacs, superscriptaddress, citeautoscript, onecolumn, nofootinbib]{revtex4-2}
\usepackage{lineno,hyperref}
\usepackage{graphicx}
\usepackage{amssymb}
\usepackage{amsmath}
\usepackage{adjustbox}

\begin{document}

\title{Linewidths and energy shifts of the electron-impurity resonant states in quantum wells with infinite barriers}

\author{Pavel~A.~Belov}
\email{pavelbelov@gmail.com}
\affiliation{Spin Optics Laboratory, St. Petersburg State University, Ulyanovskaya 1, 198504 St. Petersburg, Russia}

\date{\today}

\begin{abstract}
The linewidths and the energy shifts of the resonant states of the impurity electron in GaAs-based quantum wells (QWs) with infinite barriers are calculated. The two-dimensional Schr\"{o}dinger equation for the charge impurity in the QW is solved by the developed finite-difference method combined with the complex-scaling technique. A dependence of the linewidths and energy shifts on the QW width for the impurity localized in the center of QW is studied. The calculated results extend and improve theoretical estimations of these quantities in the GaAs-based QW by Monozon and Schmelcher [Phys. Rev. B \textbf{71}, 085302 (2005)]. In particular, in agreement with their studies we obtain that resonant states originating from the second quantum-confinement subband have negligibly small linewidths. In contrast to previous estimations, we show that for the QW widths of the order of the impurity's Bohr radius the linewidths of resonant states associated to the third quantum-confinement subband linearly depend on the thickness of the QW. We show how the previous theoretical predictions can be improved for such QW widths. We also calculate linewidths for the case when the electron impurity is localized away from the center of the QW.
\end{abstract}
\maketitle
%
%
\section{Introduction}

Many studies have been devoted to the charge impurities and the bound electron-hole pairs, excitons, in semiconductor quantum wells (QWs)~\cite{bibLutt,Bassani,Bald1971}.
The early experimental works have mainly addressed the fundamental properties of these quasiparticles, namely the lifetime and the radiative characteristics as well as the ways to tune them~\cite{Goebel,MatsusuePRL,Deveaud,Bishop}.
The renewed interest in these structures is based on the certain applications, in particular, on the quantum-cascade lasers~\cite{Kazarinov,Faist,Chow} as the primary sources of the Terahertz radiation~\cite{Alferov,Koehler,Firsov,Wienold,Fujita,MakhovJLum}.

From the theoretical point of view, charge carriers and excitons in heterostructures with QWs are interesting model systems, allowing for an accurate theoretical treatment and comparison with experiment.
Since the first works of Bastard on impurities~\cite{BastardChargeImpurity1981,BastardChargeImpurity1984} and following numerous works on excitons in QWs~\cite{Bastard3,Thoai,Ivchenko,Kavokin2005,Shuvayev,Haug,Nazmitdinov,BCS,Khramtsov,Loginov}, the variational calculations of bound state energies have become a standard approach.
Most of the variational studies were devoted to the particle ground state.
The point is that the simplest choice of the trial function in the variational method (see for example Refs.~\cite{Greene,Thoai,Khramtsov}) allows one to reliably simulate only the ground state.
To calculate the energy levels of the excited states, for example originating from upper quantum-confinement subbands in QW, one needs the specially chosen set of trial functions, that complicates numerical solution.

Although most of the studies addressed the systems with real parameters of QWs,
some works investigated the bound states in very narrow QWs, when the Coulomb potential can be effectively treated as a two-dimensional one~\cite{Portnoi,Duclos,BelovPhysE}.
Another simplification is that the Coulomb coupling of the bound states at upper subbands with the continuum of lower subbands was ignored.
In fact, this coupling leads to the resonant (quasibound) states in the continuum.
They are characterized not only by the energy levels, but also by their nonzero linewidths.

The resonant states of the charge impurities were investigated both analytically and numerically in several works~\cite{Priester,Yet,Bloom2003,Aleshkin2004,MONOZON,Aleshkin2008}.
The most instructive one is Ref.~\cite{MONOZON}, in which Monozon and Schmelcher considered resonances in
the model of three quantum-confinement subbands in narrow QW with infinite barriers.
This allowed them to derive the analytical expressions for energy shifts and linewidths of the resonant states of an impurity electron as well as of an electron-hole pair in a QW of width $L$.
In particular, they obtained that the linewidth scales with $L$ as $L^{4}$,
that is in agreement with the Fano theory of resonances~\cite{Fano,Mir}.
However, this result is applicable only for very narrow QWs, when the QW thickness is much smaller than the Bohr radius $a_{B}$: $L\ll a_{B}$.
For the well-known bulk GaAs, the Bohr radius $a_{B}=\epsilon \hbar^2/(m_{e} e^2)$ of the electron impurity is of about $10$~nm.
Thus, for GaAs-based QWs such a condition is hardly be applicable, in particular because for widths of $1$-$2$~nm the envelope function approximation is not accurate enough.
For GaAs, the effective mass method is appropriate for the thicknesses of QWs $L>3$~nm~\cite{Belov2017}.

In the present paper, we numerically study the linewidths of the resonant (quasibound) states of the electron impurity in a single QW with infinite barriers.

The upper quantum-confinement subbands are coupled to the unrestricted in-plane electron-impurity motion of lower subbands
and lead to a resonant nature of the upper states.
The resonant states are well known in quantum physics~\cite{Feshbach,Fano,Friedrich}:
the linewidth $\hbar\Gamma$ of the energy level, $E$, determines the lifetime of a resonant state.
In semiconductor physics, the quasibound states of the electron-hole pairs in QWs are in the scope of intensive studies~\cite{BCS,Trifonov2015,Grig,Sadreev}.
There are also works devoted to the observation and studies of their bound states in the continuum~\cite{Hsu,Rivera} characterized by very small nonradiative broadenings~\cite{Stillinger}.
The electron-hole pairs in QWs are effectively three-body systems~\cite{Kez,BY}, i.e. they are more complicated objects than the charge impurities.
Therefore, before studying the features of electron-hole resonances, one has to investigate
resonant states of the charge impurity.

In order to calculate the linewidths, we use the complex-scaling technique~\cite{Moiseyev}.
It transforms the scattering problem into the boundary value problem with zero boundary conditions.
It was rigorously established for the description of resonance characteristics in the early seventies~\cite{Aguilar,Balslev,Simon,SimonECS}.
Since then, many quantum scattering problems in nuclear and atomic physics have been studied by this technique~\cite{Weinhold,Ho,Kukulin,Elander,Vinitsky,Telnov,Myo}.
In semiconductor physics,
a feasibility of complex-scaling calculations of the electron-hole nonradiative linewidths was firstly demonstrated in Ref.~\cite{Belov2019}.
Recently, the complex scaling has also been applied to identify energies and linewidths of the resonant series of the electron-hole pairs in bulk cuprous oxide~\cite{Rommel2020,RommelPRB,RommelPRB2021}.

We study the resonance characteristics using the developed numerical algorithm combined with the complex-scaling rotation.
The bulk GaAs material parameters are used in calculations~\cite{Vurgaftman}.
Our numerical algorithm is based on the finite-difference discretization of the two-dimensional Schr\"{o}dinger equation.
This approach has already allowed us to obtain precise exciton energies for a wide range of QW widths and different potential profiles~\cite{Khramtsov,Grig,BelovPhysE}.
In the current work, we readily determine the spectrum of the bound and resonant states of the electron impurity localized in the center of the GaAs-based QW, using the complex-scaling technique we distinguish the proper states from the artificially discretized continuum, and classify the states over subbands.
A dependence of the linewidths of resonant states on the QW width and the index of the quantum-confinement subband is studied.
In particular, we obtain that the resonant states associated to the second quantum-confinement subband have negligibly small linewidths. Thus, they have very long lifetime and can be considered as an example of the bound states in the continuum.
We extend the results presented by Monozon and Schmelcher for very narrow QWs to wider ones by
providing new calculated data on the linewidths of resonant states of the third quantum-confinement subband for QWs widths up to 100~nm.
We also show that in order to obtain more precise linewidths for QW widths of order of the Bohr radius
one can use the complete formulas derived in Ref.~\cite{MONOZON} without approximations,
which the authors, then, employed to achieve the final analytical expressions of Ref.~\cite{MONOZON}.
Additionally, we calculate linewidths for the case when the electron impurity is localized away from the center of the QW.

\section{Theory}

\subsection{The energy levels of the charge impurity in QW}

We consider an electron impurity (a donor center) localized at the distance $b$ from the center of QW.
The $z$ axis is chosen along the growth axis and $\vec{\rho}$ is the radius-vector in the QW plane. The point $z=0$ is in the center of QW.
Within the effective mass approximation, the wave function of the electron impurity at the position $\mathbf{r}(\vec{\rho},z)$ satisfies the Schr\"{o}dinger equation~\cite{MONOZON}
\begin{equation}
\label{eq1}
\left(-\frac{\hbar^{2}}{2m_{e}} \Delta-\frac{e^{2}}{\epsilon \sqrt{|\vec{\rho}|^{2}+(z-b)^{2}}} + V_{e}(z) \right) \Psi(\mathbf{r}) = E \Psi(\mathbf{r}).
\end{equation}
Here, $\Delta$ is the three-dimensional Laplace operator, $m_{e}$ is the electron mass, $e^{2}$ is the square of the electron charge, and $\epsilon$ is the dielectric permittivity constant.
The potential $V_{e}(z)$ is defined as
\begin{equation}
\label{eqV}
V_{e}(z) = \left\{
  \begin{array}{lr}
    0 & \mbox{ if  } |z|<L/2 \\
    \infty & \mbox{ if  } |z| \ge L/2
  \end{array}
\right. .
\end{equation}
To simplify Eq.~(\ref{eq1}), we introduce the polar coordinates $\vec{\rho}=(\rho,\phi)$ within the QW plane.
Separating the angle variable and taking into account only cylindrically symmetrical solutions, Eq.~(\ref{eq1}) is reduced to the two-dimensional equation
\begin{equation}
\label{eq2}
\left(
-\frac{\hbar^{2}}{2m_{e}} \frac{\partial^{2}}{\partial z^{2}} + V_{e}(z)
-\frac{e^{2}}{\epsilon \sqrt{\rho^{2}+(z-b)^{2}}}
-\frac{\hbar^{2}}{2m_{e}}
\left[\frac{\partial^{2}}{\partial \rho^{2}} - \frac{1}{\rho} \frac{\partial}{\partial \rho} + \frac{1}{\rho^{2}} \right]
 \right) \chi(\rho,z) = E \chi(\rho,z).
\end{equation}
The function $\chi(\rho,z)$ is related to the radial part of the wave function $\Psi(\rho,z)$ as $\chi(\rho,z)=\rho \, \Psi(\rho,z)$.

\begin{figure}[htbp!]%
\begin{center}
\includegraphics*[width=0.4\linewidth]{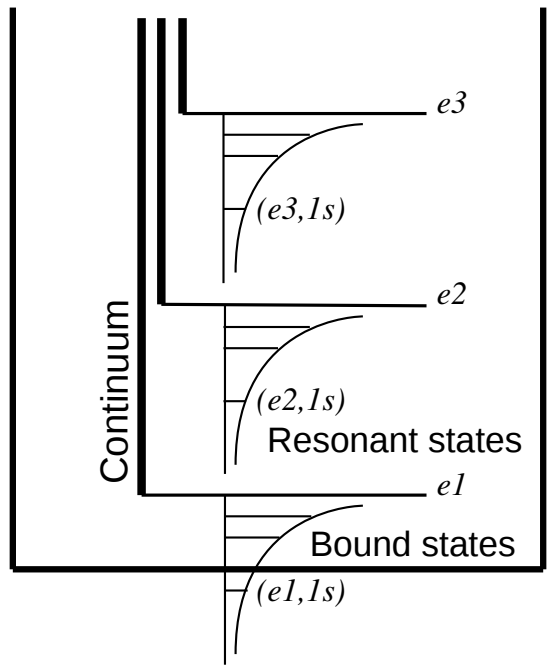}
\caption{\label{WNQW}Scheme of the energy levels of the impurity electron in a QW with infinite barriers. The quantum-confinement subbands $ej$, $j=1,2,3$, as well as the corresponding branches of the continuum are shown. The bound and resonant states are explicitly denoted.}
\end{center}
\end{figure}

The energy levels of the electron impurity in QW are governed by Eq.~(\ref{eq2}).
The scheme of energy levels can be obtained in a similar way as it was done in Ref.~\cite{BelovPhysE} for the exciton in QW.
This spectrum is shown in Fig.~\ref{WNQW}.
In the figure, the quantum-confinement subbands, originating from a quantization in QW, are shown by long horizontal lines.
They are defnied as~\cite{Landau}
\begin{equation}
\label{E3DZ}
E_{ej}=\frac{\hbar^{2}}{2m_{e}} \left( \frac{\pi j}{L} \right)^{2}.
\end{equation}
Corresponding quantum-confinement wave functions are~\cite{Ivchenko}
\begin{equation}
\label{wfQC}
\psi_{j}(z)= \sqrt{\frac{2}{L}} \left\{
  \begin{array}{lr}
    \cos(j\pi z/L) & \text{ if } j=1,3,5,\ldots\\
    \sin(j\pi z/L) & \text{ if } j=2,4,6,\ldots
  \end{array}
\right. .
\end{equation}
Below each quantum-confinement subband, there is a set of Coulomb-like energy levels which can be approximated for the very narrow QWs as energies of the 2D Coulomb potential~\cite{Duclos,BelovPhysE}
\begin{equation}
\label{ECoulomb}
E_{N}^{C} = -\frac{4 \, \text{Ry}}{(2N-1)^{2}}.
\end{equation}
Here $\text{Ry}=m_{e}e^4/(2\epsilon \hbar^2)$ is the Rydberg energy of the impurity electron in bulk GaAs.

According to Ref.~\cite{BelovPhysE}, we classify the energy levels by the index $j=1,2,3,\ldots$ of the quantum-confinement subband as well as by the principal quantum number $N=1,2,3,\ldots$ of the Coulomb-like level: $(ej,Ns)$. For example, the lowest energy level is $(e1,1s)$, see Fig.~\ref{WNQW}.

Below the lowest quantum-confinement subband, $e1$, there are energy levels corresponding to the square-integrable solutions.
They are associated with the discrete part of the spectrum.
The continuous part of the spectrum takes place above the subband $e1$.
Energies of the continuum correspond to the square-nonintegrable, scattering solutions of Eq.~(\ref{eq2}), i.e. solutions which propagate far away from the interaction domain.

The Coulomb-like energy levels of upper quantum-confinement subbands are coupled to the continuum of the lower subbands.
These states are the resonant (quasibound) states. They are characterized not only by the energy, but also by the linewidth broadening, determining the lifetime of a resonant state.
The time evolution of resonant (quasibound) states can be expressed as $\exp\left[-i(E-i\hbar\Gamma) t/\hbar \right]$.
Therefore, energies $E$ and lifetimes $\tau=\hbar/(2\hbar\Gamma)$ of these states can be defined by complex values with the negative imaginary part $\mathrm{Im}[E]=-\hbar\Gamma$.
Here, $\Gamma>0$ is the half-width at half-maximum (HWHM) linewidth broadening of a resonance~\cite{Ivchenko,Khramtsov}.
It is related to the full-width at half-maximum (FWHM) definition~\cite{Landau} of broadening as $\Gamma_{\mathrm{FWHM}}=2\Gamma$.

\subsection{Analytical results on the energy shifts and linewidth broadenings}
Due to the Coulomb coupling of the upper subbands with the lower ones as well as with the continuum of lower subbands one obtaines the shifts of energy levels and the linewidth broadenings, respectively.
In Ref.~\cite{MONOZON}, these quantities were obtained analytically based on the model of the two and three quantum-confinement subbands in narrow QW with infinite barriers.
In a framework of the two-subband model Monozon and Schmelcher obtained the following energies of the Coulomb-like series associated to the second subband $e2$. In our notation, they are
\begin{equation}
\label{E2}
E_{e2,Ns} = E_{e2}+E_{N}^{C}+\Delta E_{e2,Ns} - i\hbar\Gamma_{e2,Ns},
\end{equation}
where
\begin{equation}
\label{E2dE}
\Delta E_{e2,Ns} = \frac{8 \, \text{Ry}}{(2N-1)^{3}} \left( \frac{L}{a_{B}} \right)
\end{equation}
and
\begin{equation}
\label{E2g2}
\hbar\Gamma_{e2,Ns}=0.
\end{equation}
Here, we fixed $b=0$ to significantly simplify the analytical formulas. We outline the original expressions of Ref.~\cite{MONOZON} in the Appendix.
The first two terms in Eq.~(\ref{E2}) refer to the model of a very narrow QW, when the Coulomb potential becomes effectively two-dimensional, $\sim-\rho^{-1}$.
The energy shift $\Delta E_{ej,Ns}$ is caused by the effect of the three-dimensional Coulomb potential $\sim(\rho^{2}+z^{2})^{-1/2}$, compared to that from the two-dimensional one.
The zero linewidth $\hbar\Gamma_{e2,Ns}$ originates from the different parity of the wave function of the subband $e2$ with respect to $e1$ (and $e3$). The matrix element of the Coulomb potential calculated for these states is exactly zero.
Meanwhile, for $b\ne 0$ such a degeneracy is lifted and $\hbar\Gamma_{e2,Ns}$ becomes nonzero.

The energies of the Coulomb-like set of the third quantum-confinement subband $e3$ are 
\begin{equation}
\label{E3}
E_{e3,Ns} = E_{e3}+E_{N}^{C}+\Delta E_{e3,Ns} - i\hbar\Gamma_{e3,Ns},
\end{equation}
where
\begin{equation}
\label{E3dE}
\Delta E_{e3,Ns} = \left( 1- \frac{4}{9 \pi^2} \right) \left[ 1+ \frac{4}{\pi^4} \left( \frac{L}{a_{B}} \right)^{2} \right] \frac{8 \, \text{Ry}}{(2N-1)^{3}} \left( \frac{L}{a_{B}} \right)
\end{equation}
and
\begin{equation}
\label{E3g3}
\hbar\Gamma_{e3,Ns} = \left( 1- \frac{4}{9 \pi^2} \right) \left[ 1- \frac{4}{\pi^2} \right] \frac{8 \, \text{Ry}}{\pi^3 (2N-1)^{3}} \left( \frac{L}{a_{B}} \right)^{4}.
\end{equation}
Thus, the linewidths of the energy levels associated to the third subband are nonzero. The quantum-confinement wave function of this subband is of the same symmetry as that of the first one.
In fact, the linewidth $\hbar\Gamma_{e3,Ns}$ is appeared to be proportional to $L^{4}$, that is in accordance with the Fano theory~\cite{MONOZON,Fano}.

It is worth noting that Eqs.~(\ref{E2}-\ref{E3g3}) are obtained using the following approximations: 1) a few quantum-confinement subbands are taken into account; 2) very narrow QWs, when $L \ll a_{B}$, are considered.
In the present work, we study QWs of various widths. We obtain the energy shifts and linewidths numerically without mentioned approximations. Moreover, we consider the original Eq.~(\ref{eq2}), that means that many quantum-confinement subbands are taken into account.

\subsection{Complex-scaling technique}

For simple textbook models, e.g. one-dimensional QWs, the resonance positions and, in particular, the linewidths can be determined analytically relatively easy~\cite{Landau,Moiseyev,Rapedius}.
For more complicated systems, the analytical derivation of resonance characteristics is less straightforward (see, for example, Ref.~\cite{MONOZON} to realize the faced problems).
The main point is that, unlike the bound states,
the quasibound (resonant) ones
are not represented by the square-integrable solutions of the Schr\"{o}dinger equation.
As a result, their energies are beyond the discrete spectrum of the Hamiltonian and the convential variational methods become inappropriate for treatment of scattering solutions.

The reliable method for calculation of resonance broadenings, the complex-scaling technique, has been established in Refs.~\cite{Aguilar,Balslev,Simon,SimonECS} in the seventies.
Since then, many quantum scattering problems in nuclear and atomic physics have been studied by this technique~\cite{Weinhold,Ho,Kukulin,Moiseyev,Elander,yakovlev,Vinitsky,Telnov,Myo,belovFB}.
It states that the complex energies of resonant states can be identified in the discrete spectrum of the non-Hermitian Hamiltonian $H_{\theta}=H(r \, \text{exp}(i\theta))$, obtained from the initial one $H(r)$ by scaling of the coordinate as
\begin{equation}
\label{rotation}
r\to r \, \text{exp}(i\theta).
\end{equation}
Here $\theta>0$ is the scaling angle, i.e. the angle of a rotation of the coordinates into the complex upper half-plane.
Such a scaling allows one to associate resonant states
with the discrete spectrum by making the outgoing scattering waves $\sim\text{exp}(i\sqrt{E}\,r)$ as $r\to\infty$ to be square-integrable $\sim\text{exp}(i\sqrt{E}\,r \cos\theta-\sqrt{E}\,r \sin\theta)$ by the appropriate choice of the angle $\theta$.
As a result, the continuous spectrum becomes rotated into the complex lower half-plane by the double angle $2\theta$. The bound states remain unchanged for arbitrary $\theta$ and one can distinguish them from the discretized continuum.
If the angle of the rotation $\theta$ is large enough, then the unknown complex energies also appear in the sector of the double angle $2\theta$.
If they are independent of the angle of the rotation, they correspond to the resonance positions.
\begin{figure}[htbp!]
\begin{center}
\includegraphics*[width=0.5\linewidth]{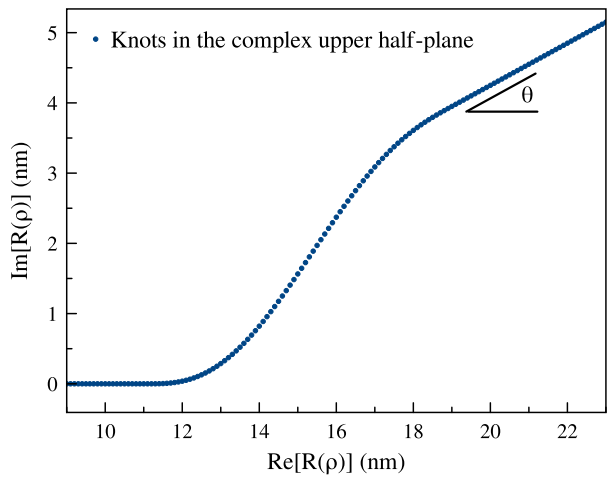}
\caption{Knots of the rotated contour for the variable $\rho$. The smooth exterior complex-scaling rotation is shown. The smooth curve starts at $\rho_{0}=11$~nm and finishes at $\rho_{1}=19$~nm.}
\label{figCS}
\end{center}
\end{figure}

Fig.~\ref{figCS} shows an example of the complex scaling of variable $\rho$ of Eq.~(\ref{eq2}) into the upper complex half-plane.
The figure shows a so-called smooth exterior rotation~\cite{Rescigno,Elander,yakovlev,belovFB}.
It implies a change of the variable $\rho$ to the complex function $R(\rho)$ according to the formula
\begin{equation*}
\label{e090}
R(\rho) = \left\{
  \begin{array}{l l}
    \rho & \text{ if } \quad \rho<\rho_{0}\\
    \rho_{0}+s(\rho;\theta,\rho_{0},\rho_{1}) & \text{ if } \quad  \rho_{0} \le \rho < \rho_{1} \\
    \rho_{0}+(\rho-\rho_{0}) \exp{i\theta} & \text{ if } \quad \rho_{1} \le \rho
  \end{array} \right. .
\end{equation*}
In this case, the limiting behavior of $R(\rho)$ as $\rho \to \infty$ is as it would be for the sharp rotation defined by Eq.~(\ref{rotation}).
However, comparing to the sharp rotation, for the exterior one the derivatives of $R(\rho)$ over $\rho$ in the vicinity of the point $\rho=0$ are zero.
Moreover, we additionally introduced the smooth complex-valued function $s(\rho;\theta,\rho_{0},\rho_{1})$ which
matches two limiting behaviors at the interfaces $\rho=\rho_{0}$ and $\rho=\rho_{1}$ and guarantees appropriate smoothness.

\subsection{\label{Psec3} Numerical method}

For the calculation of the linewidths, $\hbar\Gamma$, we implemented the complex scaling into our finite-difference method of the numerical solution of Eq.~(\ref{eq2}).
Since our method has already been described in detail in Refs.~\cite{Khramtsov,BelovPhysE},
we only briefly mention the key points.
In the current case, we use the second-order finite-difference discretization on the equidistant grids over each of two variables of complex-scaled Eq.~(\ref{eq2}).
It leads to an eigenvalue problem with the sparse block-tridiagonal non-Hermitian matrix~\cite{Korneev,Nugumanov}.
Several lowest eigenvalues of this matrix are calculated by the Arnoldi algorithm~\cite{Sorensen}.
As a result, we were able to compute the energies and the linewidths of the resonances in QWs of various widths.
We use the material parameters for bulk GaAs which are given in Refs.~\cite{BelovPhysE,Vurgaftman}: $m_{e}=0.067~m_{0}$ and $\epsilon=12.53$.

\section{Results}

\subsection{Dependence of the linewidths on the angle of the rotation}

\begin{figure}[htbp!]%
\begin{center}
    \begin{minipage}[t]{.45\textwidth}
        \centering
        \includegraphics[width=\textwidth]{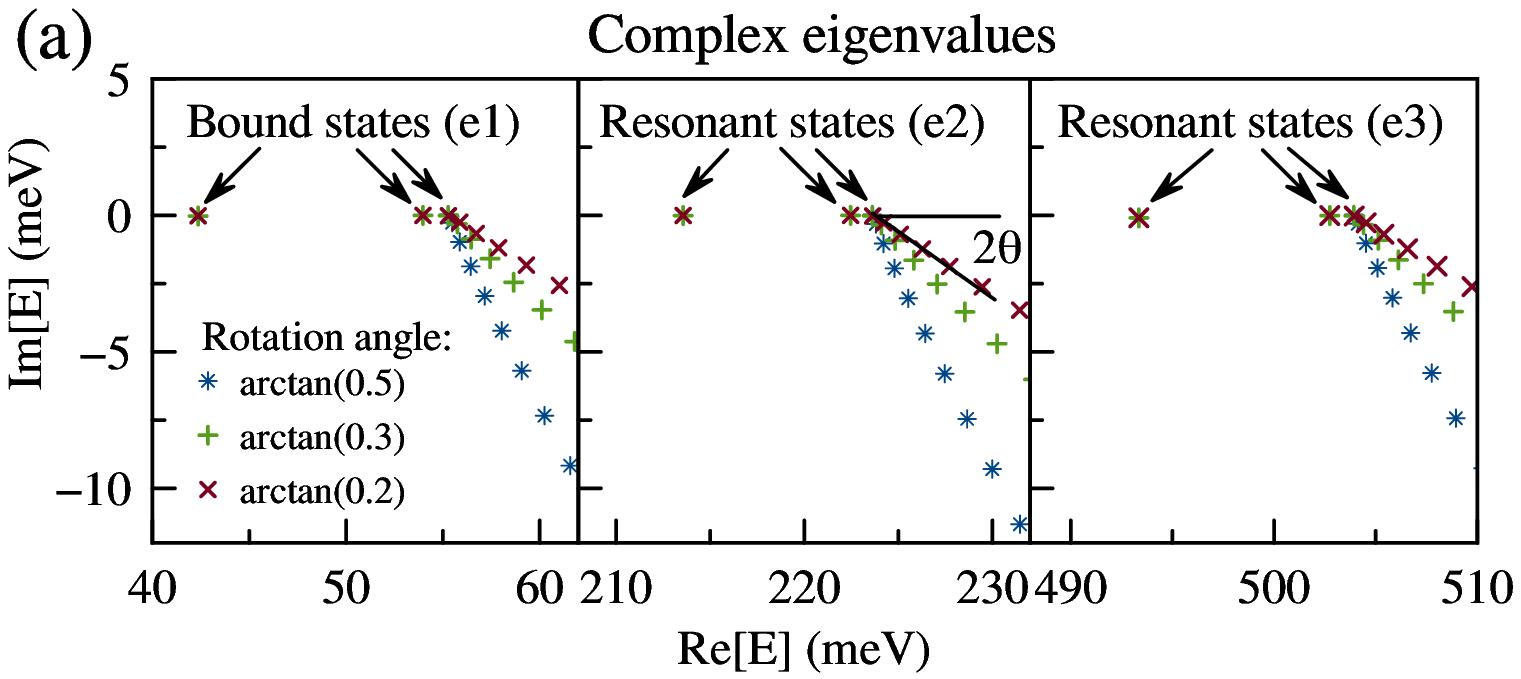}
    \end{minipage}
    \hfill
    \begin{minipage}[t]{.45\textwidth}
        \centering
        \includegraphics[width=\textwidth]{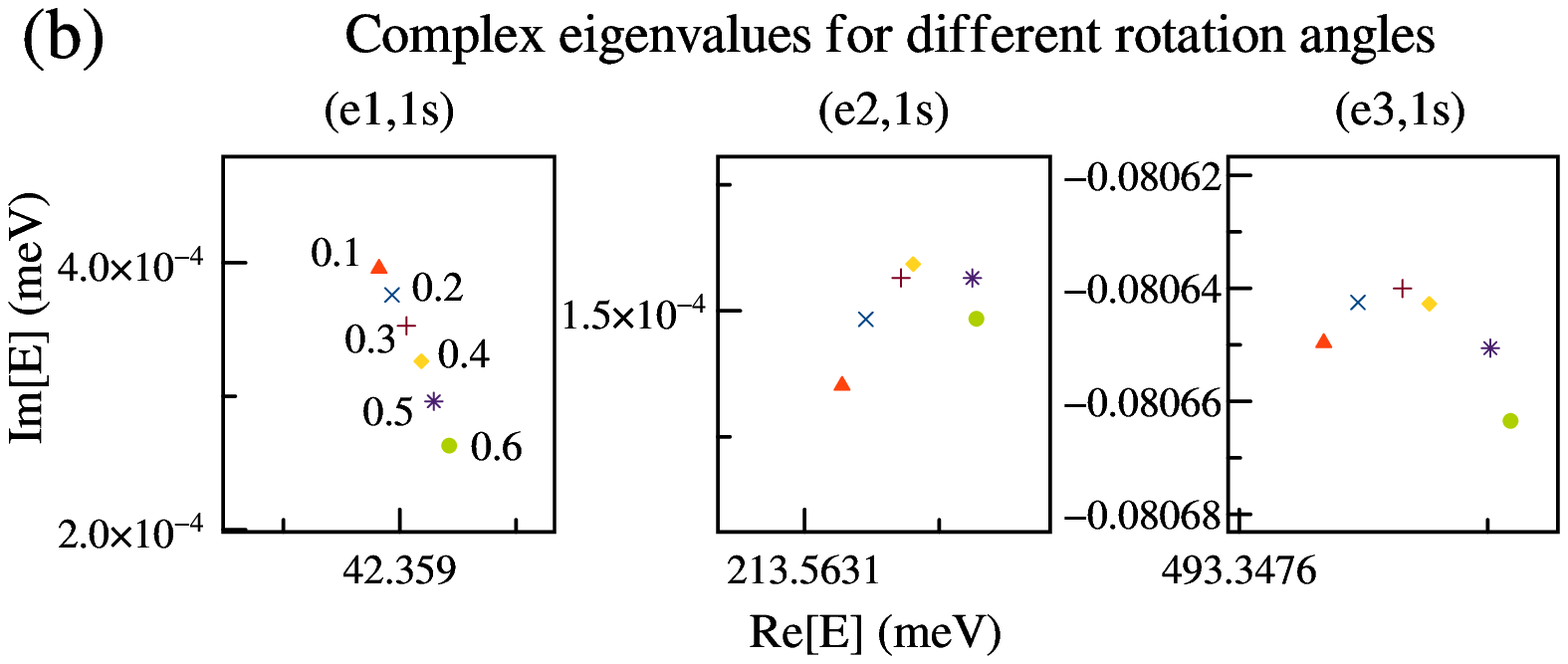}
    \end{minipage}
\caption{\textit{(a):} Calculated eigenvalues of the complex-scaled Eq.~(\ref{eq2}), defining the states of the electron impurity in the infinite-barrier QW of thickness $L=10$~nm. Three different angles of the rotation are demonstrated. The left panel shows eigenvalues corresponding to the lowest quantum-confinement level, $e1$, i.e. corresponding to the bound states. The central panel shows the eigenvalues of the energy level $e2$, i.e. the resonant states. The right panel shows the eigenvalues of the quantum-confinement enegy level $e3$. The discretized branches of the rotated continuum also shown. \textit{(b):} Detailed plots of the calculated eigenvalues of the complex-scaled Eq.~(\ref{eq2}), defining the electron impurity in the $10$-nm QW with infinite barriers. Tangents of the rotation angles $\theta$ are also denoted nearby to the corresponding points. The relatively small positive imaginary parts of the complex eigenvalues associated to the subbands $e1$ and $e2$ are due to the uncertainty of the numerical method.}
\label{figEIGCS}
\end{center}
\end{figure}

Using our numerical method, we calculated several lowest eigenvalues of the complex-scaled Eq.~(\ref{eq2}) with $b=0$ for different QW widths.
Only the smooth exterior rotation of the variable $\rho$ was performed.
It started at $\rho_{0}=11$~nm and finished at $\rho_{1}=19$~nm by approaching the straight line inclined by some angle $\theta$ to the real axis.
For example, the numerical results for the QW width $L=10$~nm are shown in Fig.~\ref{figEIGCS}~(a) by three panels for better visibility. Each panel corresponds to one set of Coulomb-like energies below the certain quantum-confinement subband $ej$, $j=1,2,3$.
One can see the branches of the discretized continuum rotated by the double angle $2\theta$. The complex eigenvalues, independent of $\theta$ and corresponding to the bound and resonant states, are explicitly denoted.

The positions of resonances do not depend on the angle of the rotation, $\theta$, if the numerical scheme is exact.
In fact, due to uncertainty of the numerical method, our calculated series of eigenvalues slightly depend on the rotation angle.
To study this dependence, we calculated the complex eigenvalues $E-i\hbar\Gamma$ as functions of $\theta$.
They are shown in detail in Fig.~\ref{figEIGCS}~(b).
One can see that there is indeed a weak dependence of the results on the angle of the rotation.
Moreover, since the bound states should have zero linewidths, we can estimate the accuracy of our calculations as of about $4\times 10^{-4}$~meV.
%

\begin{figure}[htbp!]%
\begin{center}
    \begin{minipage}[t]{.45\textwidth}
        \centering
        \includegraphics[width=\textwidth]{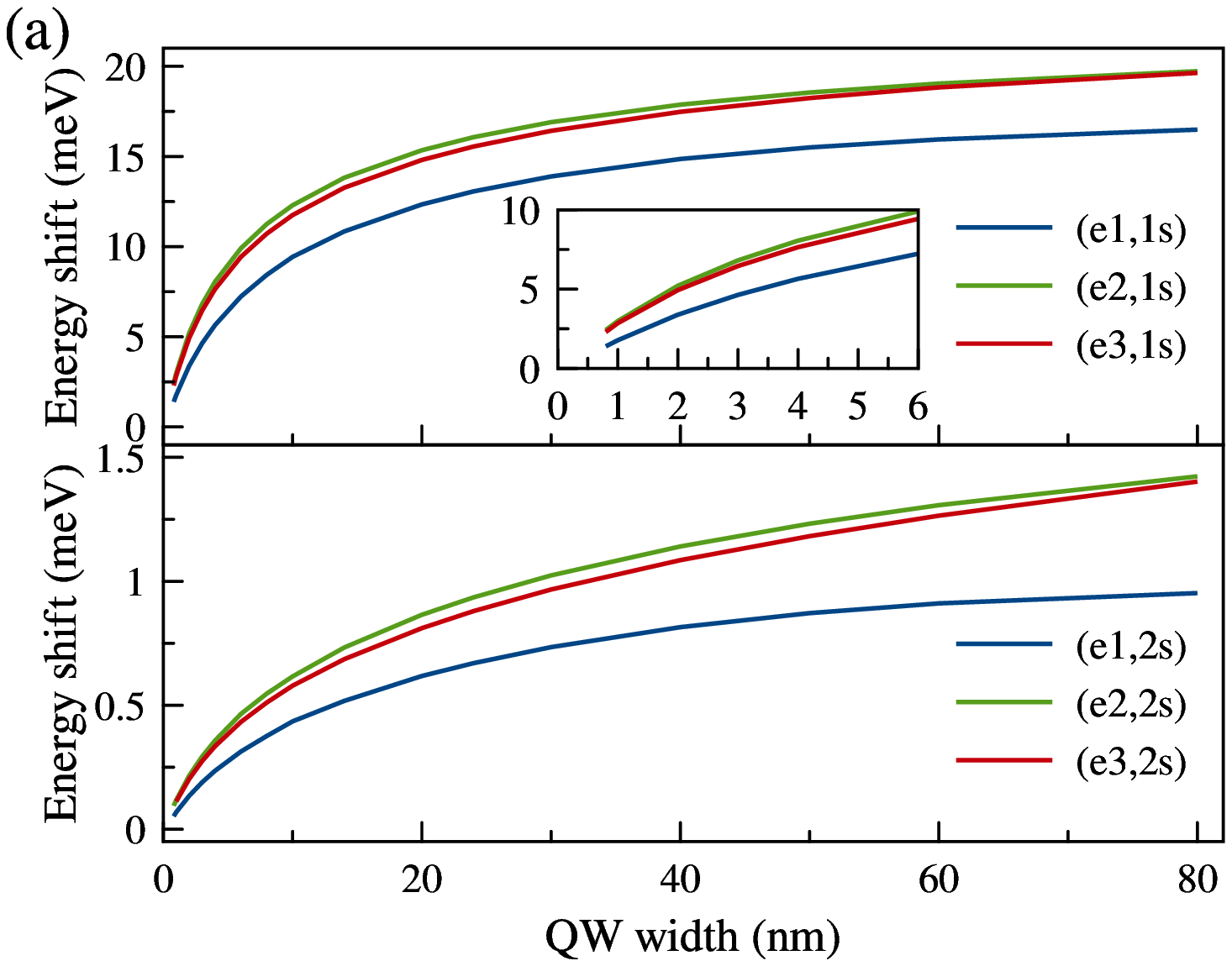}
    \end{minipage}
    \hfill
    \begin{minipage}[t]{.45\textwidth}
        \centering
        \includegraphics[width=\textwidth]{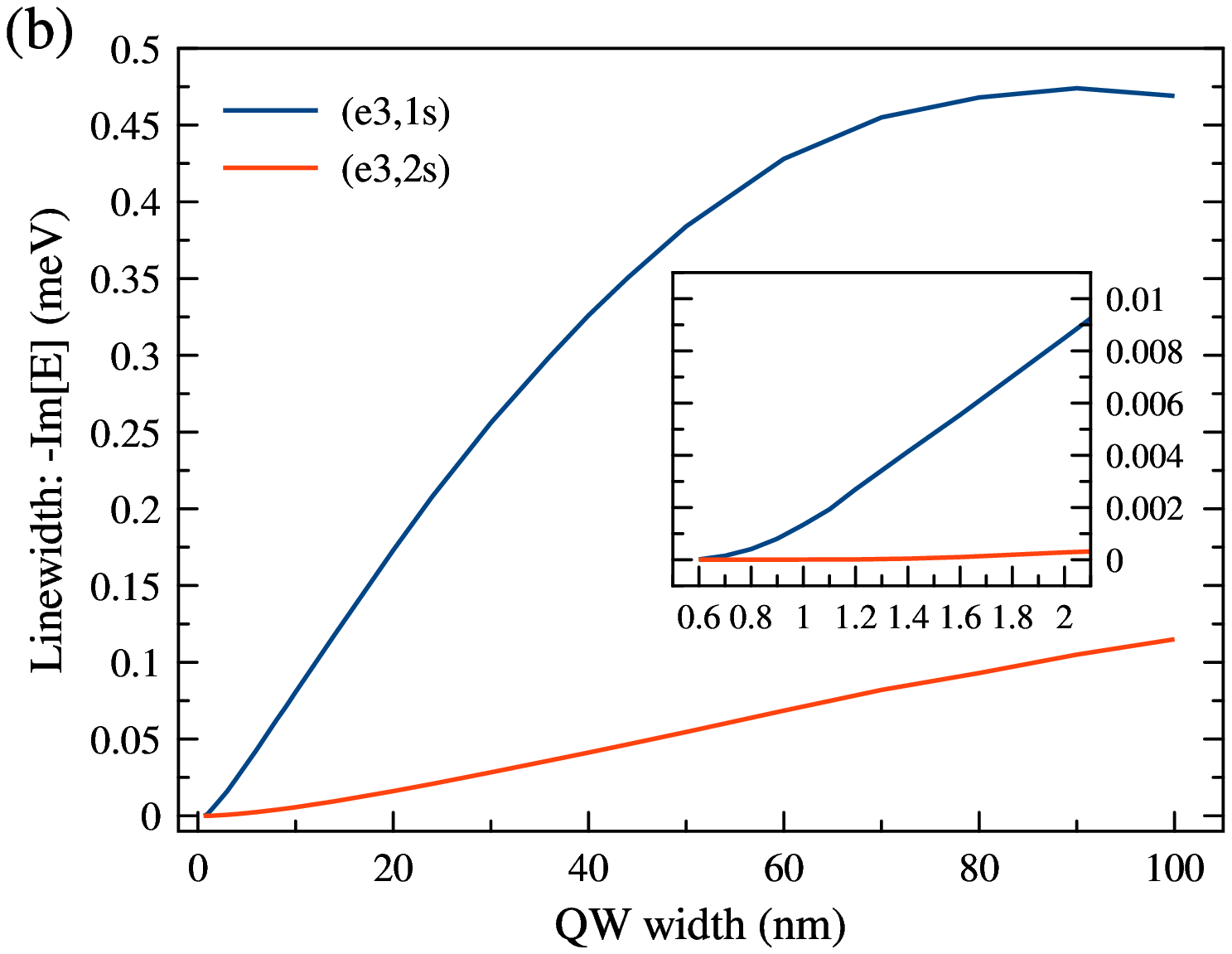}
    \end{minipage}
\caption{\textit{(a):} Calculated energy shifts of the resonant states as well as of the bound states as a function of QW width. The upper (lower) panel shows the data for $1s$-like ($2s$-like) states. The states of the subband $e1$ correspond to the bound states, whereas ones of subbands $e2$ and $e3$ correspond to the resonant ones. \textit{(b):} Calculated linewidths of the resonant states $(e3,1s)$ and $(e3,2s)$ as a function of QW width $L$.}
\label{eEnergyShiftLB100}
\end{center}
\end{figure}

\subsection{Dependence of the energy shift and the linewidth broadening on the thickness of the QW}

We calculated the energies of the electron impurity in the infinite-barrier QW of widths $L=0.6-100$~nm.
To keep the symmetry of the system, we localized the impurity electron in the center of QW ($b=0$).
%
%
Comparing calculated energies with the energies $E_{ej}+E_{N}^{C}$ for $j=1,2,3$ and $N=1,2$, we found the shifts of the energy levels due to the Coulomb coupling.
The obtained energy shifts, highlighting the difference of the impurity energy from that in the model of very narrow QW, are shown for bound and resonant states in Fig.~\ref{eEnergyShiftLB100}~(a).
One can see that energy shifts grow with an increase of the width of QW.
Moreover, for narrow QWs the energy shifts grow faster than for wide QWs.
Such a saturation exhibits the fact that the rapid change of energies in the model of 2D Coulomb ($\sim \rho^{-1}$)
takes place for narrower QWs, immediately after allowing the electron impurity to move over $z$-axis.
For wide QWs, the system is broad enough.
Thus, further expansion over $z$-axis weakly affects the energies.
%
%
We calculated not only the energies but also the linewidths of the resonant states.
Interestingly enough, the widths of the resonant states associated to the second quantum-confinememt subband, $e2$, appeared to be of the order of the uncertainty of calculations for the whole range of studied QW widths.
For example, the complex energy of the state $(e2,1s)$ of order of $10^{-4}$~meV is shown in the central panel of Fig.~\ref{figEIGCS} (b).
Therefore, we can numerically confirm the result of Ref.~\cite{MONOZON} that magnitudes of $\hbar\Gamma_{e2,Ns}$ are negligibly small.
Since these states are effectively uncoupled to the continuous spectrum of lower subbands, they can be regarded as the bound states in the continuum~\cite{Stillinger,Hsu}.
By the same reason, they should be sharper in the experimental spectra than other resonant states.

The linewidths of the resonant states associated to the third quantum-confinememt subband, $e3$, are evidently nonzero.
These quantities for $1s$- and $2s$-like states of this subband are shown in Fig.~\ref{eEnergyShiftLB100}~(b).
One can see almost the linear growth of $\hbar\Gamma$ for the $1s$-like state for QW widths $L=1.2-50$~nm as the QW width increases.
Then, it flattens out and saturates at about $0.5$~meV for 90~nm.
Thus, in agreement with experiments the states corresponding to the subband $e3$ should have more pronounced peaks in the measured spectra for narrower QWs~\cite{Pearah,Khramtsov,Loginov}.
For wide QWs, the measured peaks should be significantly broadened.
The linewidth of the $2s$-like state is by one order of magnitude smaller than that for the $1s$-like state. It increases for the whole range of studied QW widths.

\subsection{Comparison with the analytical results}
%
%
The results by Monozon and Schmelcher~\cite{MONOZON} predict that for very narrow QWs, when $L\ll a_{B}$, the linewidth scales with $L$ as $L^{4}$.
In our calculations, we can see such a behavior only for very small thicknesses $L=0.6-1.2$~nm, see the inset in Fig.~\ref{eEnergyShiftLB100}~(b).
Since for the electron impurity in bulk GaAs $a_{B}\approx 10$~nm, this range of $L$ indeed satisfies the above-mentioned criterium. 
In fact, for these values of $L$ the power-law fit of calculated data gives the dependence $L^{3.5\pm0.2}$ which is similar to the theoretical predictions.
\begin{figure}[htbp!]%
\begin{center}
\includegraphics*[width=0.5\linewidth]{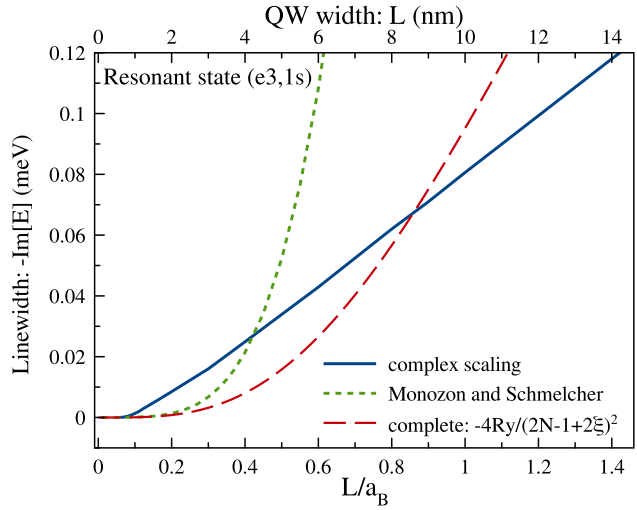}
\caption{Calculated linewidths of the resonant state $(e3,1s)$ as a function of $L/a_{B}$. Results of our calculations using the complex-scaling technique are shown by the solid line. The linewidths plotted from the final analytical expression of Ref.~\cite{MONOZON} are denoted by the dotted curve. The linewidths plotted from the intermediate complete expressions of Ref.~\cite{MONOZON} are shown by the dashed line.}
\label{FigCompare}
\end{center}
\end{figure}

Fig.~\ref{FigCompare} shows our numerical results for $(e3,1s)$ resonant state together with the theoretical predictions made by Monozon and Schmelcher as well as calculations made using the intermediate complete expressions in Ref.~\cite{MONOZON}.
Both theoretical and numerical estimations reasonably agree for $L\ll a_{B}$.
However, our numerical data and the values based on the final theoretical results by Monozon and Schmelcher significantly differ for $L\sim a_{B}$.
Such a discrepancy can arise from several approximations employed in Ref.~\cite{MONOZON} to derive the final results.
In particular, the authors approximate the complete expression
for the resonance complex energy
$E_{e3,Ns}=-4 \text{Ry}/(2N-1+2\xi)^{2}$
by the perturbation series
\begin{equation}
\label{eq1expansion}
E_{e3,Ns}\approx -\frac{4\text{Ry}}{(2N-1)^{2}}+\frac{16\text{Ry} \xi}{(2N-1)^{3}}+\frac{16\text{Ry} \xi^{2}}{(2N-1)^{4}}.
\end{equation}
with a small parameter $\xi$, which is given as the ratio of perturbation series in powers of $L/a_{B}$, see the details in the Appendix.
In the leading order, $\xi$ depends linearly on $L$, therefore for $L\ll a_{B}$ the shift of energy
and the corresponding complex parameter $\xi$ in Eq.~(\ref{eq1expansion}) are assumed to be small.
This allowed the authors to take into account only the linear term in Eq.~(\ref{eq1expansion})
as well as a few lower orders of $L/a_{B}$ in the perturbation series in both part of the ratio for $\xi$.
These approximations led finally to the analytical results for the energy shift~(\ref{E3dE}) and the linewidth~(\ref{E3g3}).

Nevertheless, the complete expressions can be used directly without the above-mentioned approximations.
It appears that the complete expressions give values [see the dashed curve in Fig.~\ref{FigCompare}] which are much closer to the ones calculated by the complex scaling technique.
One can see that both the results differ by no more than $0.02$~meV up until $L\sim a_{B}$, whereas
for such a range the originally predicted $L^{4}$-dependence from Ref.~\cite{MONOZON} exceeds our numerical data by the order of magnitude.

Interestingly enough,
the theoretical predictions in Ref.~\cite{MONOZON} evidently do not show a linear dependence on $L/a_{B}$,
whereas the calculations based on the complete expressions show a linear-like behavior for $L\sim a_{B}$.
Our numerical results are linear for even smaller $L$, already for $L\sim 0.1 \; a_{B}$.
We can claim that taking into account higher orders of $L/a_{B}$ in the perturbation series for both parts of the ratio for $\xi$ lead to more pronounced linear dependence on $L/a_{B}$.
However, the reasons of such a fact require further indepth theoretical studies.
Anyway, our numerical calculations do not rely on the perturbation theory and provide accurate results for the whole range of QWs widths.

\begin{figure}[htb]
    \begin{minipage}[t]{.45\textwidth}
        \centering
        \includegraphics[width=\textwidth]{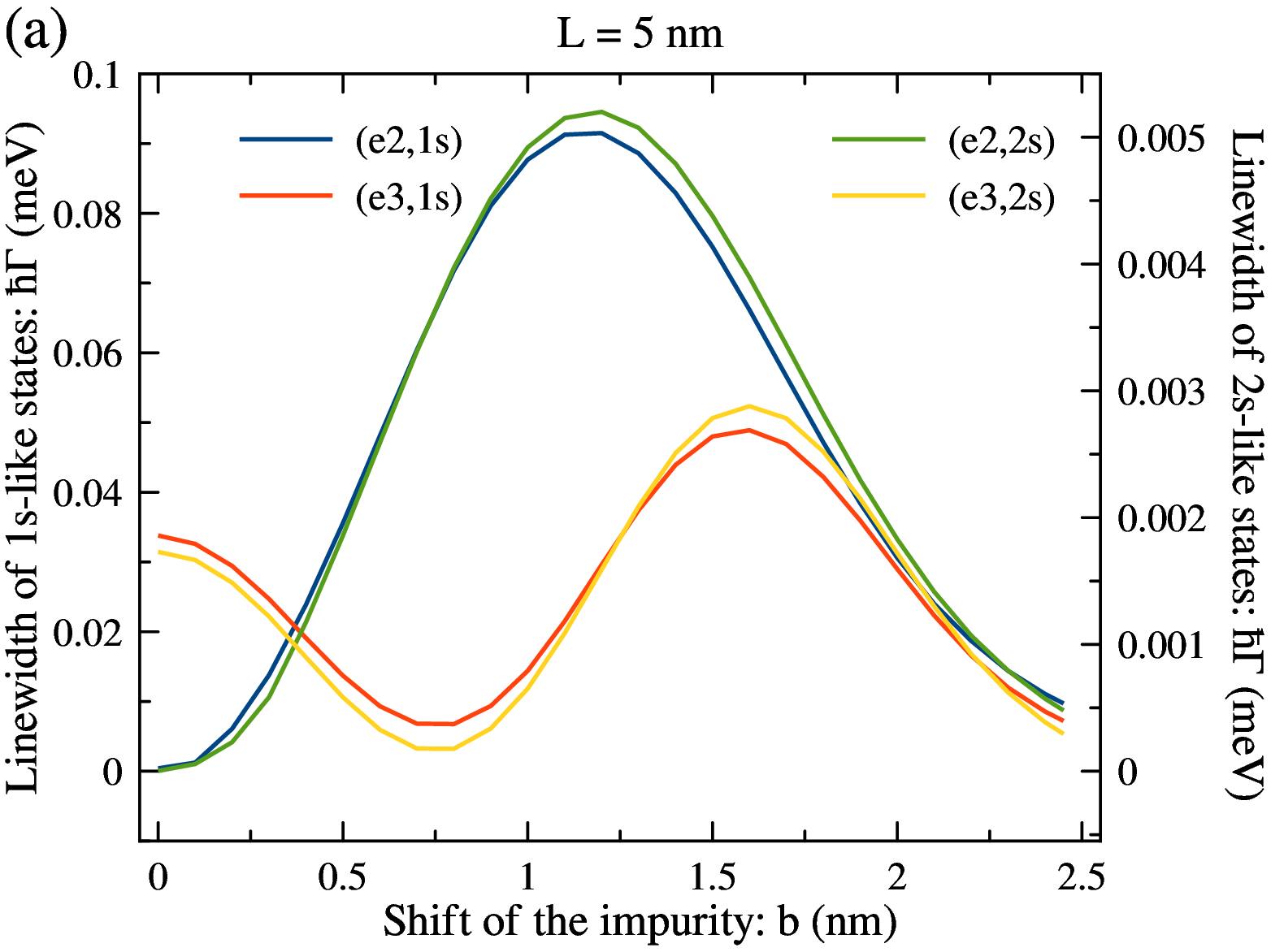}
    \end{minipage}
    \hfill
    \begin{minipage}[t]{.45\textwidth}
        \centering
        \includegraphics[width=\textwidth]{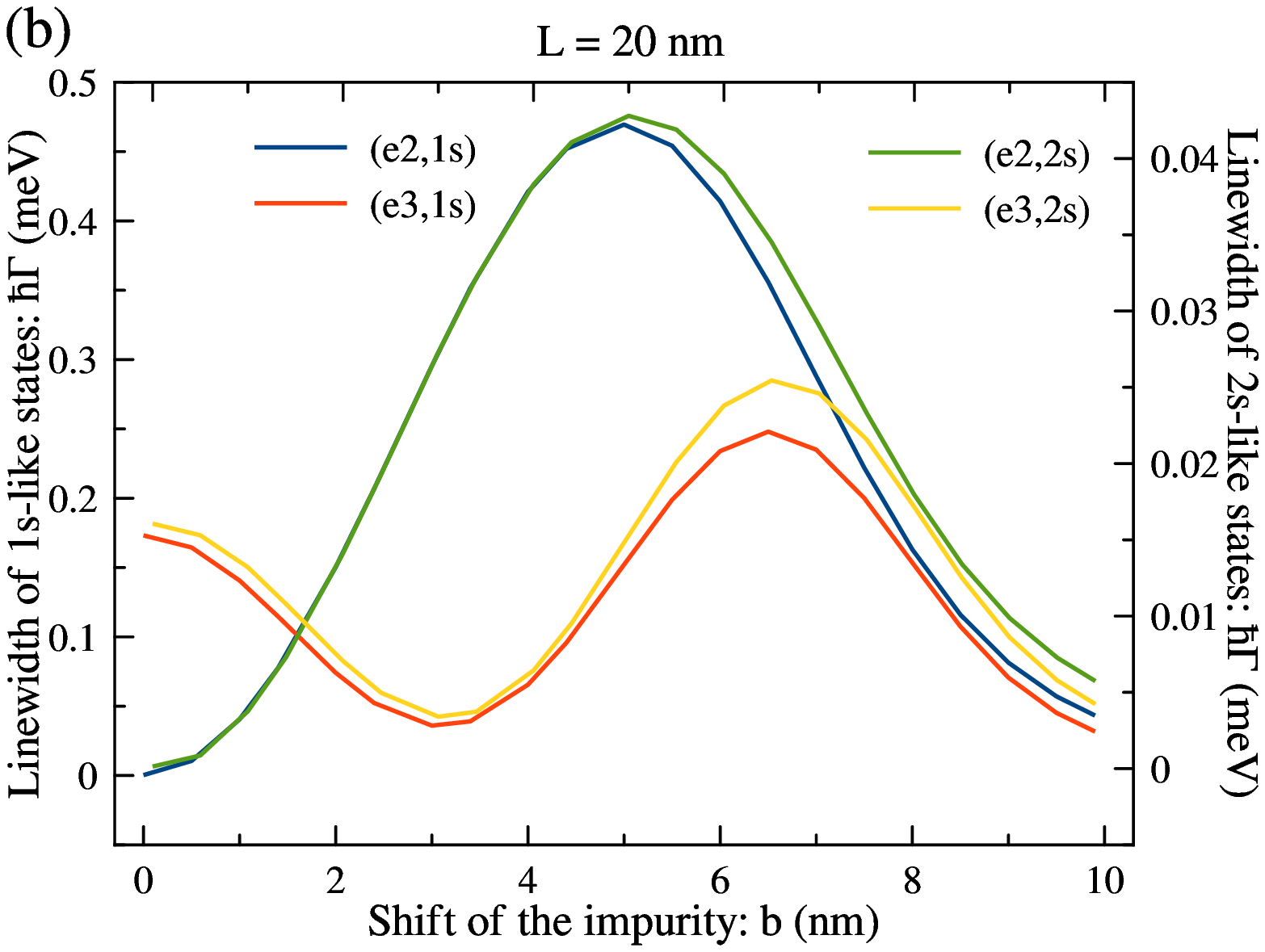}
    \end{minipage}  
    \caption{\label{FigQW5and20nm}Calculated linewidths of the resonant states belonging to the second and the third quantum-confinement subbands as functions of the shift $b$ of the impurity from the center of QW. The plot $(a)$ shows data for QW width $L=5$~nm, whereas the plot $(b)$ shows results for $L=20$~nm.}
\end{figure}

\subsection{Linewidths of the resonant states when the impurity is away from the center of the QW}

We additionally studied the case when the electron impurity is away from the center of QW, i. e. $b \ne 0$.
In such a case, a symmetry of the system is broken and, in particular,
the linewidths $\hbar\Gamma_{e2,Ns}$ [Eq.~(\ref{E2g2})] of resonant states associated to the second quantum-confinement subband become nonzero.
To show this, we calculated the linewidths as functions of the shift from the center, $b$, for QW width $L=5$~nm and $L=20$~nm.
These QW widths address the situations when the QW is two times smaller than the impurity's Bohr radius and when it is two times larger, respectively.
The obtained dependecies are shown in Fig.~\ref{FigQW5and20nm}.
The forms of the dependencies corresponding to the same states are quite similar for two studied values of $L$ and differ only by magnitudes.
The magnitudes of these dependencies for $1s$- and $2s$-like states are denoted in the plots by the left scale and by the right scale, respectively. In each plot, the dependecies for $1s$- and $2s$-like have also the similar forms.

One can see that for small $b \sim 0.1$~nm the linewidths $\Gamma_{e2,Ns}$ for $N=1,2$ grow with the increase of $b$.
For larger values of $b$, the dependencies are nonlinear.
There are maxima as well as minima of the linewidth brodenings. Thus, by varying $b$ one can achieve the situation when the state associated to the subband $e2$ has the maximal linewidth or when that one belonging to the subband $e3$ has the minimal one.
The common feature of both plots is that the linewidths become small, when the impurity is localized near the barrier.
As it is shown in Ref.~\cite{MONOZON}, the calculated linewidths are related to the matrix elements of the Coulomb potential.
This reduction of $\hbar\Gamma_{ej,Ns}$ orininates from the shift of the Coulomb potential close to the barrier, i.e. the domain where values of the quantum-confinement wave functions in the matrix element are relatively small. Thus, the coupling of states also becomes small.

\section{Conclusions and outlook}
In summary, we calculated the energies and the linewidths, $\hbar\Gamma$, of resonant states of the impurity electron in a single QW with infinite barriers.
These resonant states originate from the Coulomb coupling of upper quantum-confinement subbands with the continuum of the lower subbands.
The two-dimensional Schr\"{o}dinger equation describing the impurity electron states was solved by the finite-difference discretization method combined with the complex-scaling technique.
A dependence of linewidth broadenings and energy shifts on the QW width and the index of the quantum-confinement subband was studied.
We obtained that the resonant states corresponding to the second quantum-confinement subband have negligibly small linewidths.
Thus, they can be regarded as examples of the bound states in the continuum.
The linewidths of states associated to the third quantum-confinement subband are nonzero.
For $L \le 1$~nm, the calculated values of the linewidth scale with $L$ approximately as $ L^{4}$, reaching
at $L=1.2$~nm the value of about $0.003$~meV, that is in agreement with results obtained by Monozon and Schmelcher in Ref.~\cite{MONOZON}.
For QW widths $L=1.2-50$~nm, the linewidth broadens linearly with an increase of $L$, then it flattens out and saturates at about $0.5$~meV for $L\sim 100$~nm.
We extended the results obtained in Ref.~\cite{MONOZON} for very narrow QWs, $L\ll a_{B}$ to the QW of the thickness of order of the impurity's Bohr radius $a_{B}$ and larger by providing new numerical data on $\hbar\Gamma$ of resonant states of the third quantum-confinement subband.
We also showed how one can apply the complete expressions derived in Ref.~\cite{MONOZON} without approximations to obtain more precise magnitudes of the linewidths of the resonant states for QW widths of order of the Bohr radius.
Additionally, we calculated the linewidths for the case when the electron impurity is localized away from the center of the QW.

The numerical methods of this paper can be applied to the model of the excitons in cuprous oxide QWs~\cite{Zelinska}.
The obtained results are also important for contemporary experimental studies of Cu$_{2}$O-based heterostructures~\cite{Giessen}.
The estimations and the dependencies of the linewidth broadenings on the QW width provide valuable information for spectroscopy measurements of these structures.

\section*{Acknowledgments}
The author is grateful to Prof. J\"{o}rg Main and his research group for the stimulating discussions and careful reading of the manuscript.
Financial support from RSF (grant No. 19-72-20039) is acknowledged.
Calculations were carried out using the facilities of the Resource Center ``Computer Center of SPbU''.

\section*{Appendix}
Here, we outline the theoretical results of Ref.~\cite{MONOZON} and show in detail how they can be improved for $L \sim a_{B}$ if one calculates the linewidths using the complete expressions in Ref.~\cite{MONOZON} instead of taking the lowest-order approximations to obtain the analytical results.

In Ref.~\cite{MONOZON}, the model of three quantum-confinement subbands in QW with infinite barriers was solved.
Within this model, the complex energies of the electron impurity are defined by the parameter $\xi$ via Eq.~(\ref{eq1expansion}).
For $L\ll a_{B}$, this parameter is assumed to be small enough to approximate the energies by the constant and the linear in $\xi$ terms, see the right-hand side of Eq.~(\ref{eq1expansion}).
Monozon and Schmelcher found that $\xi$ can, in turn, be given by the ratio of perturbation series in the small parameter $L/a_{B}$:
\begin{equation}
\label{appendEq2}
\xi=\frac{g_{33}}{2}\frac{1+\frac{i\pi}{4}\left(g_{11}+g_{22}\right)-\frac{\pi^{2}}{16} g_{11} g_{22}}{1+\frac{i\pi}{4}\left(g_{11}+g_{22}\right)-\frac{1}{4} \left( g_{31}^{2} + g_{32}^{2} \right)-\frac{i\pi}{16}\left( g_{22} g_{31}^{2} + g_{11} g_{32}^{2} \right)-\frac{\pi^{2}}{16} g_{11} g_{22}}.
\end{equation}
%
Here, the integrals are given as
$$
g_{jk}=\left\langle \psi_{j}(z) \left| \frac{4|z-b|}{a_{B}} \right| \psi_{k}(z) \right\rangle.
$$
For QW with infinite barriers, the integrals include the well-known wave functions~(\ref{wfQC}) and are evaluated exactly:
\begin{align*}
g_{11} &= \left(\frac{L}{a_{B}} \right) \left[ 1+\frac{4}{\pi^{2}} \left( (t/2)^{2} - \cos^{2}{(t/2)}  \right) \right], \\
g_{22} &= \left(\frac{L}{a_{B}} \right) \left[ 1+\frac{1}{\pi^{2}} \left( t^{2} - \sin^{2}{t} \right) \right], \\
g_{33} &= \left(\frac{L}{a_{B}} \right) \left[ 1+\frac{4}{9\pi^{2}} \left( (3t/2)^{2} - \cos^{2}{(3t/2)}  \right) \right], \\
g_{32} &= \left(\frac{L}{a_{B}} \right) \frac{1}{\pi^{2}} \left[ \frac{8}{25} \sin{(5t/2)} - 8 \sin{(t/2)} \right], \\
g_{31} &= \left(\frac{L}{a_{B}} \right) \frac{4}{\pi^{2}} \cos^{4}(t/2), \\
\end{align*}
where $t=2\pi b/L$.
For $L\ll a_{B}$, the series in Eq.~(\ref{appendEq2}) can be truncated leaving only the leading orders of $L/a_{B}$ which are necessary to obtain the unknown quantity.
By doing likewise, as well as by using the linear approximation in Eq.~(\ref{eq1expansion}), Monozon and Schmelcher obtained following approximations for
the real part of the complex energy (the enery shift)
\begin{eqnarray}
\nonumber \Delta E_{\mathrm{Re}}=\frac{8 \text{Ry}}{(2N-1)^{3}} g_{33} \left[ 1+ \frac{1}{4} \left( g_{31}^{2}+g_{32}^{2} \right) \right]=
\frac{8 \text{Ry}}{(2N-1)^{3}} \left(\frac{L}{a_{B}} \right) \left[ 1+\frac{4}{9\pi^{2}} \left( (3t/2)^{2} - \cos^{2}{(3t/2)}  \right) \right] \times \\
\times \left\{1+\frac{1}{4} \left(\frac{L}{a_{B}} \right)^{2} \left[\frac{16}{\pi^4} \cos^{8}(t/2) + \frac{1}{\pi^4} \left( \frac{8}{25} \sin{(5t/2)} - 8 \sin{(t/2)} \right)^{2} \right]  \right\}
\end{eqnarray}
and for the imaginary part (the linewidth)\footnote{Please note the two times difference with $\Gamma$ in Ref.~\cite{MONOZON} due to the definition of the complex energy as $E-i\Gamma/2$ there.}
\begin{eqnarray}
\nonumber \Delta E_{\mathrm{Im}}=-\frac{i \pi \text{Ry}}{2(2N-1)^{3}} g_{33} \left[ g_{11} g_{31}^{2} + g_{22} g_{32}^{2} \right]
\nonumber =-\frac{i}{\pi^{3}}\frac{8 \text{Ry}}{(2N-1)^{3}} \left(\frac{L}{a_{B}} \right)^{4} \left[ 1+\frac{4}{9\pi^{2}} \left( (3t/2)^{2} - \cos^{2}{(3t/2)}  \right) \right] \times \\
\times \left\{ \left[ 1+\frac{4}{\pi^{2}} \left( (t/2)^{2} - \cos^{2}{(t/2)}  \right) \right] \cos^{8}(t/2) +
\left[ 1+\frac{1}{\pi^{2}} \left( t^{2} - \sin^{2}{t} \right) \right] \left[ \frac{2}{25} \sin{(5t/2)} - 2\sin{(t/2)} \right]^{2}
 \right\}.
\end{eqnarray}
%
\begin{figure}[htb]
    \begin{minipage}[t]{.45\textwidth}
        \centering
        \includegraphics[width=\textwidth]{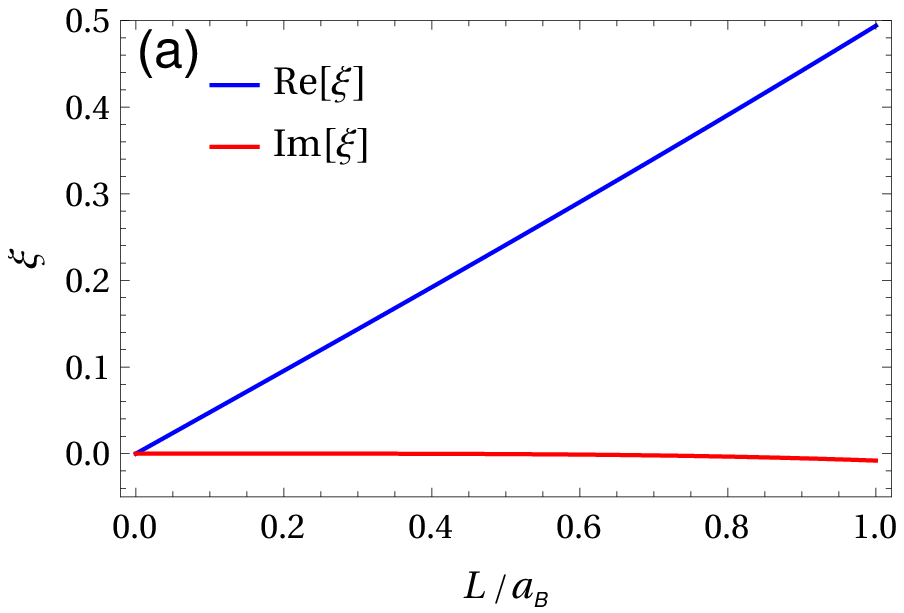}
    \end{minipage}
    \hfill
    \begin{minipage}[t]{.45\textwidth}
        \centering
        \includegraphics[width=\textwidth]{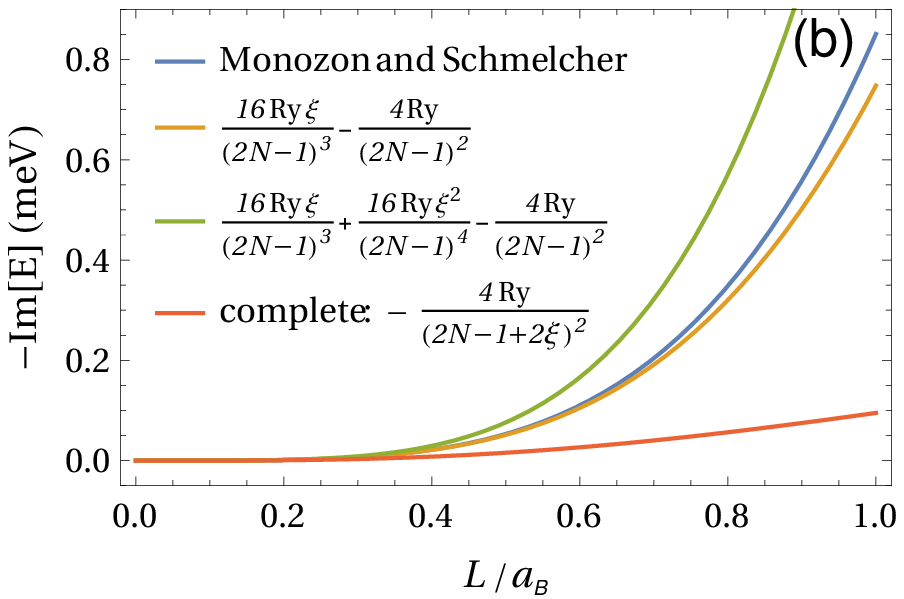}
    \end{minipage}  
    \caption{\label{xiFigappendFig1}\textit{(a):} Values of $\xi$ as a function of $L/a_{B}$. \textit{(b):} The imaginary part of the complex energy, i.e. the linewidth, calculated using the complete expressions derived in Ref.~\cite{MONOZON}, see Eqs.~(\ref{eq1expansion},\ref{appendEq2}) and different applied approximations, also including the original result by Monozon and Schmelcher.}
\end{figure}

Using all the terms in Eq.~(\ref{appendEq2}) we can evaluate $\xi$ for different ratios $L/a_{B}$.
The dependence $\xi(L/a_{B})$ is shown in Fig.~\ref{xiFigappendFig1}~(a).
One can see that already for $L/a_{B}=0.3$, the quantity $2\xi$ in the denominator of $E_{e3,Ns}=-4 \text{Ry}/(2N-1+2\xi)^{2}$ is of about $0.3$, that is not so small comparing to unity.
Simultaneously, the last term in Eq.~(\ref{eq1expansion}) starts playing a role for such $L$, because, due to the large real part, its imaginary part becomes of the order of $\text{Im}[\xi]$.
Thus, the linear in $\xi$ approximation becomes reasonably inappropriate.

We evaluate the complex energies for $L<a_{B}$ by different ways to compare.
To do it, we calculate $\xi$ using all the terms in Eq.~(\ref{appendEq2}) and use it further to calculate the complex energy by the complete equation $E_{e3,Ns}=-4 \text{Ry}/(2N-1+2\xi)^{2}$ as well as by the linear and quadratic approximations in the right-hand side of Eq.~(\ref{eq1expansion}).
The comparison of the imaginary parts, the linewidths, of the obtained values with the result by Monozon and Schmelcher is shown in Fig.~\ref{xiFigappendFig1}~(b).
For small values of $L/a_{B}$ both the shown curves coincide, however for $L\sim a_{B}$ there is a significant difference.
Using all the available terms in Eq.~(\ref{appendEq2}) to evaluate $\xi$ and calculating exactly the energy as $E_{e3,Ns}=-4 \text{Ry}/(2N-1+2\xi)^{2}$ allows us to significantly improve the approximation of the imaginary part of the complex energy for $L\sim a_{B}$.
We see in Fig.~\ref{FigCompare} that for $L\sim a_{B}$ the results obtained from the total expression for $\xi$ and the complete equation  for $E_{e3,Ns}$ give linewidths which agree much better with our numerical data.
There is also the linear-like dependence of this quantity for such values of $L$.

As a result, we see that the complete expressions for $\xi$ and $E_{e3,Ns}$ significantly improve theoretical predictions for the linewidths for $L\sim a_{B}$.

\section*{References}

\end{document}